\documentstyle[12pt]{article}
\newtheorem{Le}{Lemma}
\newtheorem{Th}{Theorem}
\newcommand{\be}{\begin{equation}}
\newcommand{\ee}{\end{equation}}

\begin{document}
\title{Finite W-algebras}
\author{ T.Tjin \thanks{email: tjin@phys.uva.nl}
\\ Instituut voor Theoretische Fysica \\
Valckenierstraat 65 \\ 1018 XE Amsterdam \\ The Netherlands }
\maketitle

\begin{abstract}
Finite versions of W-algebras are introduced by
considering (symplectic) reductions of finite dimensional
simple Lie algebras. In particular
a finite analogue of $W^{(2)}_3$ is
introduced and studied in
detail. Its unitary and non-unitary, reducible and irreducible
highest weight representations are constructed.
\end{abstract}

\section*{Introduction}
Ever since the introduction of W-algebras by Zamolodchikov
\cite{Zam} they have been the subject of intense investigation
by both physicists and mathematicians. Originally they were
introduced as extensions of conformal symmetry. Later it
was shown that (their classical counterparts) were already
implicitly known in the theory of integrable evolution equations
in 2 dimensions. The W-algebras of Zamolodchikov are the
quantization of the so called Gelfand-Dickii brackets on the
space of pseudo-differential operators \cite{B,FL}.
In \cite{DS,BFFWO}
it was shown that the Gelfand-Dickii brackets are
Hamiltonian reductions of the so called Kostant-Kirillov
bracket on the duals of Kac-Moody algebras. This established a
relation between Kac-Moody algebras and W-algebras the latter
being a reduction of the former (another relation is the coset
construction \cite{Cos,BBSS}).

In \cite{PB} a new reduction of the affine $sl_3$ KM algebra was
investigated which lead to a new W-algebra known as
$W^{(2)}_3$. It seemed that there should be a more general theory
of reductions of KM-algebras each leading to an extended conformal
algebra. In \cite{BTV} it was shown that the Drinfeld-Sokolov
reductions and the reduction leading to $W^{(2)}_3$ are special
cases of reductions associated to $sl_2$ embeddings. In the present
paper finite versions of these W-algebras are
introduced by investigating reductions of the Kostant-Kirillov
Poisson bracket of finite dimensional
simple Lie algebras instead of KM algebras.
Again it is possible to associate a reduction
to every $sl_2$ embedding. The resulting algebras are finite
versions of the W-algebras constructed in \cite{BTV}.
As a nice example we will construct
the finite analogue of $W^{(2)}_3$ and investigate
its representation theory.

In section 1 we briefly discuss (symplectic) reduction
and show how one can apply it to the Kostant-Kirillov
Poisson structure of a simple Lie algebra
to obtain finite W-algebras.
In particular the finite version
of $W_3^{(2)}$ is constructed.
In section 2 we investigate the representation theory of
this algebra. It turns out that it has unitary and non-unitary
highest weight representations. It is possible that
these representations play a similar role in the representation
theory of $W_3^{(2)}$ as do the representations of a simple
Lie algebra in the representation theory of its affinization.

\section{Reductions of finite dimensional simple Lie algebras}

In this section we explain what we mean by reduction. Our
presentation will be rather brief, however one can find more
details in the references quoted below.

Let $(M,\omega )$ be a symplectic manifold, i.e. $M$ is a
manifold and $\omega$ is a closed non-degenerate 2-form \cite{AM}.
Also let $\{\phi_i \}_{i=1}^q$ be a set of smooth functions
on M. Denote by $C$ the submanifold
\be
C=\{ p \in M \mid \phi_i(p)=0 \;\; \mbox{for all} \;\;
i=1,...,q \} \nonumber
\ee
of M, and by $j:C \rightarrow M$ the canonical embedding of
$C$ in $M$. The question one asks now is: is the manifold
$(C,j^{\ast}\omega )$ symplectic ? In general the answer to this
question is no. However as we shall see a particular coset space
of $C$ {\em is} symplectic.

In order to stay within physics terminology we call the
functions
$\phi_i$
constraints.
The problem with the form $j^{\ast} \omega$ on $C$  is not
that it is not closed, for we have
\be
d(j^{\ast} \omega ) = j^{\ast} d \omega =0 \nonumber
\ee
but that it may no longer be non-degenerate. This degeneracy is
caused by the first class constraints
(these are the constraints that, on $C$, Poisson commute with all
other constraints \cite{Dirac}). In fact the Hamiltonian
vectorfields of the first class constraints, which are tangent
to $C$, span the kernel of the 2-form $j^{\ast} \omega $.
A different way of saying this is that the first class
constraints generate gauge transformations on the constraint
surface. We can remove this degeneracy (gauge invariance)
by passing to a quotient manifold. Let us make this more precise.
The Hamiltonian vectorfields of the first class constraints
form an involutive system on the constraint manifold $C$. Therefore
by Frobenius' theorem $C$ foliates into leaves which are
exactly the integral (hyper)surfaces of these Hamiltonian
vectorfields. The tangent space at $p \in C$ of the leaf
passing through $p$ is therefore precisely the kernel
of $j^{\ast} \omega$.

It is clear that in order to  get a
symplectic manifold we have to identify the
points on a leaf. The resulting coset space is denoted by
$\bar{M}$ and is called the reduced phase space. Let
$\pi : C \rightarrow \bar{M}$ be the canonical projection,
then $\omega$ induces a symplectic
form $\bar{\omega}$ on $\bar{M}$ by
the formula \cite{AM}
\be
j^{\ast} \omega = \pi^{\ast} \bar{\omega} \nonumber
\ee
The reduced phase space has thus become a symplectic manifold.

The flows of the Hamiltonian vectorfields of the first class
constraints generate a group $H$ of transformations on $C$  that
leave the leaves invariant. The quadruple
$(C,\bar{M},H,\pi )$ is a principal fibre bundle with total
space $C$, base manifold $\bar{M}$ and structure group $H$. The
fibres are the leaves discussed above. Assuming that this
fibre bundle is trivial (as will always be the case in this
paper) we can choose a global section, i.e. a map
$s: \bar{M} \rightarrow C$ such that $\pi \circ s = id$. In
physics terminology this is called 'choosing a gauge'.
The composite map $j \circ s$ is then an embedding of
$\bar{M}$ into $M$. There is now a simple relation between
$\bar{\omega}$ and $\omega$, namely
\be     \label{AAA}
\bar{\omega}=(j \circ s)^{\ast}\omega
\ee
The easy proof goes as follows:
\be
(j \circ s)^{\ast}\omega = s^{\ast}j^{\ast}\omega =
s^{\ast}\pi^{\ast}\bar{\omega} = (\pi \circ s)^{\ast}
\bar{\omega}=\bar{\omega} \nonumber
\ee

Assume that $\bar{M}$, seen as a submanifold of $M$, can
globally be characterized by
$\phi_i(p)=0$ for $i=1,...,n$ and $p \in \bar{M}$,
where $\{\phi_i\}_{i=q+1}^{n}$ are supplementary constraints,
or 'gauge fixing conditions'. This will always be possible
in the cases we consider.

We can now give the relation between the Poisson bracket on
$M$ induced by $\omega$ and the Poisson bracket $\{.,.\}^{\ast}$
on $\bar{M}$ induced by $\bar{\omega}$. It reads
\be   \label{Dir}
\{\bar{f},\bar{g}\}^{\ast}=\overline{\{f,g\}-\sum_{ij=1}^n
\{f,\phi_i \}
\Delta^{-1}_{ij} \{\phi_j ,g \}}
\ee
where the bar means 'restrict to $\bar{M}$', and $\Delta_{ij}$
is the matrix $(\{\phi_i,\phi_j \})_{ij=1}^{n}$. The proof of
this formula, which follows directly from equation (\ref{AAA}),
will not be given here but can be found in \cite{constr}.
The bracket (\ref{Dir}) was originally discovered by
Dirac \cite{Dirac} and is therefore called the 'Dirac bracket'.
Note that the Dirac bracket can still be defined if the
Poisson structure on $M$ is not associated to a symplectic
form.

Let now $M=g$ where $g$ is a finite dimensional simple
Lie algebra. For simplicity we take $g$ to be $sl_n({\bf R})$
in this paper. On $g$ there lives the so called
Kostant-Kirillov-Poisson bracket. It is defined as follows:
let $f,g$ be smooth functions on $g$ and $J
\in g$, then
\be
\{f,g\}(J)=( J ,
[df|_J,dg|_{J} ])  \nonumber
\ee
where
$df |_{J}\in g$ is defined by
\be
\frac{d}{d \varepsilon }f(J + \varepsilon \tilde{J} )
\mid_{\varepsilon =0}=( \tilde{J} , df|_{J} ) \nonumber
\ee
and $(.,.)$ is the non-degenerate Cartan-Killing form on $g$.

In \cite{BTV} reductions of this bracket were considered in case
$g$ was an affine KM-algebra $g=\bar{g} \otimes {\bf C}[t,t^{-1}]
\oplus {\bf C}c \oplus {\bf C}d$. The constraints $\phi_i$
were determined by the branching rules of an $sl_2$ embedding
into $\bar{g}$. These reductions lead to inequivalent extended
conformal algebras for every $sl_2$ embedding.

We can construct the underlying finitely generated algebras
of
these extended conformal algebras by considering  similar
reductions of the underlying finite simple Lie algebras of the
KM algebras.

Let us consider how an $sl_2$ embedding determines the functions
$\phi_i$. \\ Let $\{T_3,T_+,T_- \} \subset g$ be an $sl_2$
subalgebra of $g$. Under the adjoint action of this
subalgebra $g$  is, in general, reducible and splits up into a direct
sum of $sl_2$
multiplets. Therefore any element $J$ of $g$ can be
written as
\be \label{negen}
J=\sum_{k=1}^{p}\sum_{m=-j_k}^{j_k} U^{k,m}(J)T_{k,m}
\ee
where $p$ denotes the total number of multiplets, $j_k$ denotes
the spin of the k$^{th}$ multiplet and $T_{k,m}\in g$ is
the grade $m$ element of this multiplet, i.e.
\begin{eqnarray}
[T_3,T_{k,m}] & = & mT_{k,m}  \nonumber \\ {}
[T_{\pm},T_{k,m}] & \sim  & T_{k,m \pm 1} \nonumber
\end{eqnarray}
We also take $T_{1,0}=T_3$, $T_{1,\pm 1}=T_{\pm}$. The quantities
$U^{k,m}$ are the coordinate functions on $g$ with respect
to the basis $\{T_{k,m}\} \subset g$ (so $U^{k,m}:
g \rightarrow {\bf R}$ is the function on $g$ assigning to $J$ its
$(k,m)$ component).

The constraints are now \cite{BTV}
\begin{eqnarray}
\phi_{k,m} & \equiv & U^{k,m} \;\; \mbox{for}\;\; m<0 \;\;
\mbox{and} \;\; k\neq 1  \nonumber \\
\phi_{1,-1} & \equiv & U^{1,-1}-1  \label{twaalf}
\end{eqnarray}
The constraints $\{\phi_{k,m} \}_{m\leq -1}$ are then all
obviously first class (there are also second class constraints
if there are $sl_2$ multiplets with half integer
spin). The (gauge) group they generate is a sub-group of $G$, the Lie
group associated to $g$. We can fix this gauge invariance
globally by bringing $J$ to the form
\be  \label{veertien}
J=\sum_{k=1}^{p}U^{k,j_k}(J) T_{k,j_k}
\ee
(this corresponds to a global section of $C$) which is called
the highest weight gauge \cite{BFFWO,BTV}. The additional (gauge
fixing) constraints are therefore
\be
\phi^{k,m} \equiv U^{k,m} \;\;\; \mbox{for}\;\; 0 \leq m < j_k
\nonumber
\ee

We can now calculate the Dirac bracket algebra w.r.t. these
constraints. This can be done directly using formula (\ref{Dir})
or using an algorithm developed in \cite{BFFWO} for the
DS reduction of the KM algebra. This algorithm can be used
for arbitrary reductions of KM algebras and can simply be
adapted to reductions of finite algebras.

As an example we consider the finite version of $W_3^{(2)}$
\cite{PB,BTV}.
The reduction leading to this algebra is associated with the
only non-principal  embedding of $sl_2$ in $sl_3$. The
8 dimensional adjoint representation of $sl_3$ splits up into
a direct sum of a spin one $(j_1=1)$, two spin one half $(j_2=
j_3=1/2)$ and one spin zero $(j_4=0)$ $sl_2$ multiplets. Writing
an arbitrary element as in eqn. (\ref{negen}) we get
\be
\left(
\begin{array}{ccc}
U^{4,0}+\frac{1}{2}U^{1,0} & U^{2,1/2} & U^{1,1}  \\
U^{3,-1/2}  & -2U^{4,0} & U^{3,1/2}  \\
U^{1,-1}    & U^{2,-1/2}   & U^{4,0}-\frac{1}{2}U^{1,0}
\end{array}
\right) \nonumber
\ee
The constraints are, according to (\ref{twaalf})
\be
U^{3,-1/2}=U^{2,-1/2}=U^{1,-1}-1=0
\ee
the last one being the only first class constraint.
The highest weight gauge (see eqn. (\ref{veertien}))
then reads
\be
\left(
\begin{array}{ccc}
U^{4,0}  &  U^{2,1/2}  &  U^{1,1}  \\
0  & -2U^{4,0} & U^{3,1/2} \\
1  &  0  &  U^{4,0}
\end{array}
\right)
\ee
Calculating the Dirac bracket algebra
of the functions $U^{4,0},U^{2,1/2},U^{3,1/2}$ and $U^{1,1}$
is then straightforward.
In
terms
of
the
(more
convenient)
generators
\begin{eqnarray}
c & = & -\frac{4}{3}(U^{1,1}+3(U^{4,0})^2) \nonumber \\
e & = & \sqrt{\frac{4}{3}} \, U^{2,1/2} \nonumber \\
f & = & \sqrt{\frac{4}{3}} \, U^{3,1/2} \nonumber  \\
h & = & -4 \, U^{4,0} \nonumber
\end{eqnarray}
it reads
\begin{eqnarray}
\{h,e \}^{\ast} & = & 2e \nonumber \\
\{h,f \}^{\ast} & = & -2f \nonumber \\
\{e,f \}^{\ast} & = & h^2+c \nonumber \\
\{c,h\}^{\ast}=\{c,e\}^{\ast} & = & \{c,f\}^{\ast}=0
\end{eqnarray}
(where restriction to $\bar{M}$ is understood).
{}From now on we will refer to this algebra,
which is the finite version of $W_3^{(2)}$, as
$\bar{W}^{(2)}_3$. The generator $c$ is the stress energy tensor
in the infinite dimensional case. We see that here it becomes
the center. This in fact holds true for an arbitrary $W_N$ algebra
(which is the reduction of the affine
$sl_N$-KM algebra associated to the
principal embedding of $sl_2$ in $sl_N$ \cite{BFFWO,DS,BTV}), i.e.
the finite  version of a $W_N$ algebra is simply
a (Poisson) abelian algebra generated by $N$ generators.
This comes as no surprise from the point of view of the coset
construction. There $W_N$ algebras are constructed as algebras of
fields of the form \cite{BBSS}
\be
d^{abc...}:J^a(z)J^b(z)J^c(z)...: \nonumber
\ee
where $d^{abc...}$ is the completely symmetric tensor of order
$\lambda_i$ $\; (i=1,...,\mbox{rank}(\bar{g}))$ of $\bar{g}$
(the underlying simple algebra of the KM algebra), so that
\be
T^{(\lambda_i)}=d^{abc...}T^aT^bT^c...     \nonumber
\ee
is the Casimir of $\bar{g}$ of order $\lambda_i$. Therefore
the finite version of a $W_N$ algebra is an algebra of Casimirs,
which is abelian by definition (it is even central if the
$W_N$ algebra is a sub-algebra of a larger one).

It is easy to work out other examples. We will not do this
here however. Instead we will consider the algebra $
\bar{W}_3^{(2)}$ just
obtained in more detail.

\section{The representation theory of
$\bar{W}_3^{(2)}$}
As a nice example of the structures that one encounters in the
study of these finite $W$-algebras we now investigate the
representation theory (on real vectorspaces)
of the commutator algebra version of
$\bar{W}^{(2)}_{3}$.

Consider the associative algebra generated
by $\{H,C,E,F\}$ subject to the relations
\begin{eqnarray}
[H,E] & = & 2E \nonumber \\ {}
[H,F] & = & -2F \nonumber \\ {}
[E,F] & = & H^2 + C \nonumber \\ {}
[H,C]=[E,C] & = & [F,C]=0 \nonumber
\end{eqnarray}
In what follows we will need the following lemma.
\begin{Le}
The following identities are true
\begin{eqnarray}
[ H,E^k ]  & = & 2kE^k \nonumber \\ {}
[ H,F^k ]  & = & -2kF^k  \nonumber \\ {}
[ E,F^k ] & = & F^{k-1} (kH^2 +kC -2k(k-1)H +
\frac{2}{3}k(k-1)(2k-1))
\end{eqnarray}
\end{Le}
Proof: One can easily
prove these identities by induction using the defining
relations of $\bar{W}^{(2)}_3$ and the formula
\be
\sum_{i=1}^{p-1}i^2 =\frac{1}{6}p(p-1)(2p-1) \nonumber
\ee
\vspace{6mm}

Consider the real span of $H$ and $C$ as the Cartan subalgebra. The
next theorem then identifies the highest weight representations
of $\bar{W}_3^{(2)}$.
\begin{Th}
Let $p$ be a positive integer and $x$ a real number.
\begin{enumerate}
\item For every pair $(p,x)$ the algebra $\bar{W}_3^{(2)}$ has	a
unique highest weight representation $W(p,x)$ of dimension p.
\item $W(p,x)$ is spanned by eigenvectors (weight vectors) of $H$.
The weights are given by
\be
\{j(p,x)-2k \}_{k=0}^{p-1}
\ee
where the highest weight $j(p,x)$ is equal to
\be
j(p,x)=p+x-1
\ee
\item The value of the central element $C$ on $W(p,x)$ is given by
\be
c(p,x)=\frac{1}{3}(1-p^2)-x^2
\ee
\end{enumerate}
\end{Th}
Proof: Define the $p$-dimensional representation $W$ as follows: Let
$v$ be a (formal) element and define
\begin{eqnarray}
H\cdot v & = & jv \nonumber \\
E\cdot v & = & 0  \nonumber \\
C\cdot v & = & cv \nonumber
\end{eqnarray}
Define the vectorspace $W$ as the span of the set
\be
\{ v_k \equiv F^k \cdot v \}_{k=0,1,2,...} \nonumber
\ee
$W$ is then a representation of $W_3^{(2)}$ and by the previous
lemma we find
\begin{eqnarray}
H\cdot v_k & = & (j-2k)v_k  \nonumber \\
E\cdot v_k & = & (k(j^2+c)-2k(k-1)j+\frac{2}{3}k(k-1)(2k-1))v_{k-1}
\nonumber \\
C\cdot v_k & = & cv_k \nonumber
\end{eqnarray}
Since we are looking for finite dimensional representations there
must be a positive integer $p$ such that $F^pv=0$. From
\begin{eqnarray}
0 & = & EF^p \cdot v \nonumber \\
  & = & (pj^2-2p(p-1)j+pc+\frac{2}{3}p(p-1)(2p-1))v_{p-1} \nonumber
\end{eqnarray}
follows that
\be
pj^2-2p(p-1)j+pc+\frac{2}{3}p(p-1)(2p-1)=0   \nonumber
\ee
Solving this equation for $j$ and demanding that it be real we find
that
\begin{eqnarray}
c(p,x) & = & \frac{1}{3}(1-p^2)-x^2  \nonumber \\
j(p,x) & = & p+x-1 \nonumber
\end{eqnarray}
where $x$ is an arbitrary real number. This proves the theorem
\vspace{6mm}.

So we get infinitely many highest weight representations of a
given dimension.
Not all these representations are irreducible. In fact there are
$(p-1)$ reducible representations of dimension $p$.
This is the
subject of the next theorem.
\begin{Th}
Let $k \in \{1, ... ,p-1\}$ then
\begin{enumerate}
\item $W(p;\frac{2}{3}k-\frac{1}{3}p)$
has a $p-k$ dimensional sub-representation isomorphic to \\
$\; W(p-k;-\frac{1}{3}(k + p))$.
\item The quotient representation is isomorphic to $W(k;\frac{2}{3}p
-\frac{1}{3}k)$, i.e.
\be
W(p;\frac{2}{3}k-\frac{1}{3}p)/W(p-k;-\frac{1}{3}(k+p)) \simeq
W(k;\frac{2}{3}p-\frac{1}{3}k)
\ee
\end{enumerate}
\end{Th}
Proof: Consider the representation $W(p,x)$. It might happen that
$E.v_k=0$ for some $k<p$. The subspace of $W(p,x)$ spanned by
$F^l.v_k$ is then an invariant subspace. Let us now consider when
this happens. Again we have to solve the equation $EF^k.v=0$. This
leads (as in the proof of the previous theorem) to the equation
\be
k(j^2+c-2(k-1)j+\frac{2}{3}(k-1)(2k-1))=0 \nonumber
\ee
This equation has three solutions for $k$, namely two we already
know
\begin{eqnarray}
k & = & 0 \nonumber \\
k & = & p \nonumber
\end{eqnarray}
and a new one
\be
k=\frac{1}{2}p+\frac{3}{2}x  \nonumber
\ee
This means that if $p$ and $x$ are such that $\frac{1}{2}p+
\frac{3}{2}x$ is an integer between 1 and $p-1$ then
\be
E.v_{\frac{1}{2}p+\frac{3}{2}x}=0 \nonumber
\ee
The sub-representation that can be built out of $v_{\frac{1}{2}p+
\frac{3}{2}x}$ is obviously a $p-k$ dimensional highest weight
representation of dimension $p-k$. It is therefore equal to
$W(p-k;\tilde{x})$ for some $\tilde{x}$. Since $C$ acts on the
sub-representation $W(p-k;\tilde{x})$ as the same multiple of
unity as on $W(p;\frac{2}{3}k-\frac{1}{3}p)$
and since $H(F^kv)=j(p;\frac{2}{3}k-\frac{1}{3}p)-2k$
we see that in order
to determine $\tilde{x}$ we have to solve the equations
\begin{eqnarray}
c(p;\frac{2}{3}k-\frac{1}{3}p) & = & c(p-k;\tilde{x}) \nonumber \\
j(p;\frac{2}{3}k-\frac{1}{3}p)-2k & = & j(p-k;\tilde{x}) \nonumber
\end{eqnarray}
Inserting the expression for $c(p,x)$ and $j(p,x)$
given in the previous
theorem we find $\tilde{x}=-\frac{1}{3}(k+p)$.
For the second part a similar argument shows that we have
to solve $\tilde{x}$ from the equations
\begin{eqnarray}
c(p;\frac{2}{3}k-\frac{1}{3}p) & = & c(k;\tilde{x}) \nonumber \\
j(p;\frac{2}{3}k-\frac{1}{3}p) & = & j(k;\tilde{x}) \nonumber
\end{eqnarray}
Again using the explicit expressions for $c$ and $j$ we find
$\tilde{x}=\frac{2}{3}p-\frac{1}{3}k$. This proves the theorem
\vspace{7mm}.

The most important representations from a physical point of
view are the unitary ones. The existence of these unitary
representations is our next subject.

Define an anti-involution $\omega$ on $W_3^{(2)}$ by
\begin{eqnarray}
\omega (E) & = & F \nonumber \\
\omega (F) & = & E \nonumber \\
\omega (H) & = & H \nonumber \\
\omega (C) & = & C
\end{eqnarray}
It is easy to check that $\omega$ is an algebra endomorphism, i.e.
that it preserves the relations of $W^{(2)}_3$.
\begin{Le}
For all positive integers $p$ and real numbers $x$ the representation
$W(p;x)$ carries a unique bilinear symmetric form $\langle .,.
\rangle$ such that
\begin{enumerate}
\item $\langle v,v \rangle =1$
\item $\langle .,.\rangle$ is contravariant w.r.t. $\omega$, i.e.
$\langle A \cdot v_l,v_k \rangle = \langle v_l,\omega (A)\cdot v_k
\rangle$ for all $A \in \bar{W}^{(2)}_3$.
\end{enumerate}
\end{Le}
Proof: Define the bilinear symmetric form $\langle.,.\rangle$
by putting
$\langle v,v \rangle =1$ and
\be
\langle v_l,v_k \rangle =\delta_{lk}\prod_{i=1}^{k}A(i)
\ee
for $l,k \in \{0,1,...,p-1\}$, where
\be
A(k)=k(j^2+c)-2k(k-1)j+\frac{2}{3}k(k-1)2k-1) \nonumber
\ee
We have to check contravariance. Obviously it suffices to
check this for the generators. For example take $A=F$, then
\be
\langle F \cdot v_l , v_k \rangle =\delta_{l+1,k}\prod_{i=1}^{k}
A(i) \nonumber
\ee
On the other hand $\langle v_l, \omega (F)\cdot v_k \rangle =
\langle v_l, E \cdot v_k \rangle$ which by lemma 1 is equal to
\begin{eqnarray}
\langle v_l,E \cdot v_k \rangle & = & \langle v_l,v_{k-1} \rangle
A(k) \nonumber \\
& = & \delta_{l+1,k}\prod_{i=1}^{k-1}A(i)A(k) \nonumber \\
& = & \delta_{l+1,k}\prod_{i=1}^{k}A(i) \nonumber
\end{eqnarray}
which proves contravariance for $F$. The proof for the other
generators is similar. Uniqueness follows by
induction on $k$ \vspace{7mm}.

So the representations $W(p;x)$ all carry bilinear symmetric
forms such that the basis vectors $v_k$ are orthogonal. This
is not enough for unitarity however for we still need to check if
the bilinear form is positive definite. This is the subject of
the next theorem.
\begin{Th}
The representation $W(p,x)$ is unitary if and only if
\be
x>\frac{1}{3}p-\frac{2}{3}
\ee
\end{Th}
Proof: We have to find out which
$W(p,x)$ have the property that the vectors $v_l$ all
have positive norm $\langle v_l , v_l \rangle >0$
for $l=1,...,p-1$. Now, by definition
\be
\langle v_l , v_l \rangle  =
\prod_{k=1}^{l} (k(j^2 +c)-2k(k-1)j+\frac{2}{3}k(k-1)(2k-1))
\nonumber
\ee
For unitarity we have to demand that this is $>0$ for all
$l=1,2,...,p-1$. This can only be the case if $A(k)>0$ for all
$k=1,...,p-1$ separately. Inserting the expressions for
$j(p;x)$ and $c(p;x)$ into this equation we find that it reduces
to
\be
x>\frac{2}{3}k-\frac{1}{3}p \nonumber
\ee
This must be true for all $k=1,...,p-1$ which is only the case if
\be
x>\frac{1}{3}p-\frac{2}{3} \nonumber
\ee
This proves the theorem \vspace{6mm}.

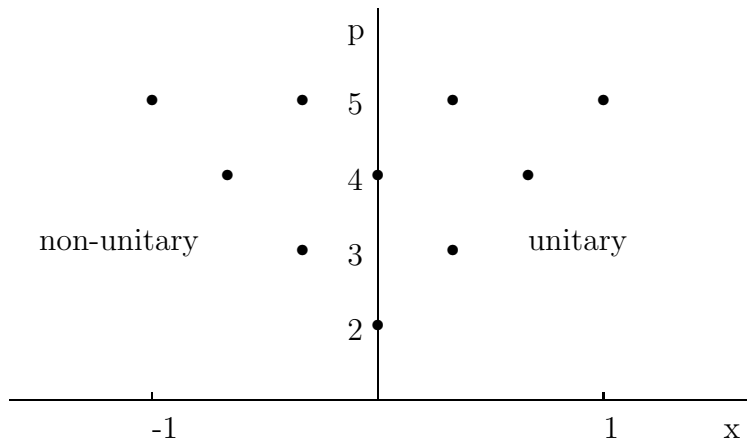
\begin{figure}[htbp]
\vspace{3.3cm}
\begin{picture}(100,55)(-81,0)
\setlength{\unitlength}{1mm}
\put (50,5){\line (0,1){52}}
\put (1,5){\line (1,0){98}}
\put (20,5){\line (0,1){1}}
\put (80,5){\line (0,1){1}}
\put (96,0){x}
\put (46,53){p}
\put (20,0){-1}
\put (80,0){1}
\put (46,13){2}
\put (46,23){3}
\put (46,33){4}
\put (46,43){5}
\put (50,15){\circle*{1.5}}
\put (40,25){\circle*{1.5}}
\put (60,25){\circle*{1.5}}
\put (30,35){\circle*{1.5}}
\put (50,35){\circle*{1.5}}
\put (70,35){\circle*{1.5}}
\put (20,45){\circle*{1.5}}
\put (40,45){\circle*{1.5}}
\put (60,45){\circle*{1.5}}
\put (80,45){\circle*{1.5}}
\put (70,25){unitary}
\put (5,25){non-unitary}
\end{picture}
\caption{The location of the reducible (depicted by dots)
and unitary representations.}
\end{figure}

Note that from theorems 2 and 3 follows that the quotient
representation of a reducible rep w.r.t. its invariant
subspace is a unitary irreducible representation.

The fact that  $\bar{W}^{(2)}_3$ has unitary representations may
be called remarkable since $sl_2({\bf R})$ does not have
unitary representations and the two algebras are so much alike.
On the other hand $\bar{W}_3^{(2)}$ is not
a deformation of $sl_2({\bf R})$ which means that it is reasonable
that its representation theory is quite different.

\section{Discussion}
In this paper we introduced the notion of finite W-algebras
and studied the specific example of $\bar{W}_3^{(2)}$ in detail.
The next step would be to apply the same analysis to the finite
W-algebras obtained by other reductions (i.e. other embeddings
in $sl_N$). In general these algebras have the following structure:
they contain $sl_M$ sub-algebras and also multiplets of generators
transforming under these $sl_M$ sub-algebras in either the
adjoint, fundamental or trivial representations. The generators
transforming in non-trivial reps should be considered as step
operators just as $E$ and $F$ in this paper. There may also be
finite  versions of $W_M$ algebras present but these
will, as we have seen earlier, be contained in the center. The
'Cartan sub-algebra' of the algebra can therefore be taken to be
the Cartan subalgebras of the $sl_M$ subalgebras together with
this center.

We have not considered tensor products of
representations of $\bar{W}^{(2)}_3$. In order to be able
to do this one should first find a suitable definition of a
tensor product representation, i.e. one needs a co-product. We
have not yet been able to construct a non-trivial coproduct
for this algebra however.

It is well known that affine Kac-Moody algebras are central
extensions of loop algebras over finite dimensional
simple Lie algebras \cite{Kac}. One wonders if the W-algebras
that are reductions of affine KM-algebras \cite{BTV} can be
constructed similarly from their finite counterparts.
Knowing such a construction could aid the development of the
representation theory of W-algebras.

Finite W-algebras should correspond to certain physical dynamical
systems of a finite number of degrees of freedom. These probably
correspond to reductions of finite Toda systems.

These points are under investigation now and will be reported on
elsewhere.

\section*{Acknowledgements}
I would like to thank Sander Bais and Peter van Driel
for pointing out the relevance of finite W-algebras to
me. I would also like to thank Koos de
Vos for useful discussions and comments.

\end{document}